\documentclass[3p,times,twocolumn]{elsarticle}
\usepackage{ecrc}
\volume{00}
\firstpage{1}
\journalname{Nuclear and Particle Physics Proceedings}
\runauth{}
\jid{nppp}
\jnltitlelogo{Nuclear and Particle Physics Proceedings}
\usepackage{amssymb}
\usepackage[figuresright]{rotating}

\begin{document}

\begin{frontmatter}

\dochead{}
\title{Recent $XYZ$ results at BESIII}
\author[label1, label2]{Lianjin WU\corref{cor1}}
\cortext[cor1]{Post-doctor}
\ead{wulj@ihep.ac.cn}
\address[label1]{Institute of High Energy Physics}
\address[label2]{19B Yuquan Road, Shijingshan District, Beijing, China}

\pagestyle{myheadings}
\markright{}

\begin{abstract}
BESIII experiment has collected about 20 fb$^{-1}$ luminosity data between $\sqrt{s}=3.8$ and 4.7 GeV via $e^{+}e^{-}$ collision.
In this talk, we present the recent $XYZ$ results at BESIII, including mass and width measurements of $Y(4220)$, search for $Y$ states with rare decay modes,
isospin parity determination of $Z_{c}^{0}(3900)$ as well as the mass and width determination, and search for new decay modes of $X(3872)$.
\end{abstract}


\end{frontmatter}

\section{Introduction}
\par The $c\bar{c}$ states below the open-charm threshold have been described excellently with potential models, 
and the theoritical and experimental sides have achieved an excellent agreement.
However, as we known, QCD allows multiquark states, such as tetra-quark, penta-quark, molecule, hybrid, glueball ... 
except for the meson (quark and antiquark) and baryon (three quarks)~\cite{Klempt:2007cp}.
Above the open-charm threshold, many states have been discovered in final states with charmonium and some light hadrons,
which are still not undertood very well due to their strange properties, and we name those states as charmoniumlike or $XYZ$ states~\cite{Brambilla:2010cs, Brambilla:2019esw}.
Also, there are many predicted states above the open-charm that have not yet been discovered~\cite{Brambilla:2010cs, Brambilla:2019esw}.
To shed light on their strange properties, those discovered or predicted states deserve us further studies or more decay mode searches.

\par $Y$ states, charmoniumlike vector states, can be produced directly via $e^{+}e^{-}$ annihilation,
while $Z$ states, isospin non-zero charmoniumlike states with heavy quark pair $c\bar{c}$ inside, and $X$ states, other states not well understood,
can be produced from $Y$ states.
BESIII experiment at BEPCII has collected about 20 fb$^{-1}$ luminosity data between $\sqrt{s}=3.8$ and 4.7 GeV via $e^{+}e^{-}$ collision,
which allows us making contribution to the $XYZ$ states spectroscopy studies. 
In this proceeding, we present the recent $XYZ$ results at BESIII.

\section{$Y$ states}

\subsection{$Y(4220)$ and $Y(4360)$}
\par In the ISR process $e^{+}e^{-}\to\gamma\pi^{+}\pi^{-}J/\psi$, BABAR observed the first $Y$ state $Y(4260)$~\cite{Choi:2003ue}. 
Later, $Y(4260)$ is comfirmed by Belle~\cite{Yuan:2007sj} and CLEO~\cite{He:2006kg}.
However, according to the fit of Born cross sections of $e^{+}e^{-}\to\pi^{+}\pi^{-}J/\psi$ at BESIII experiment~\cite{Ablikim:2016qzw},
$Y(4260)$ may consist of two components, $i.e.$, $Y(4220)$ and $Y(4320)$.

\par
\begin{figure}[hbt]
	\centering
	\includegraphics[width = 0.25\textwidth]{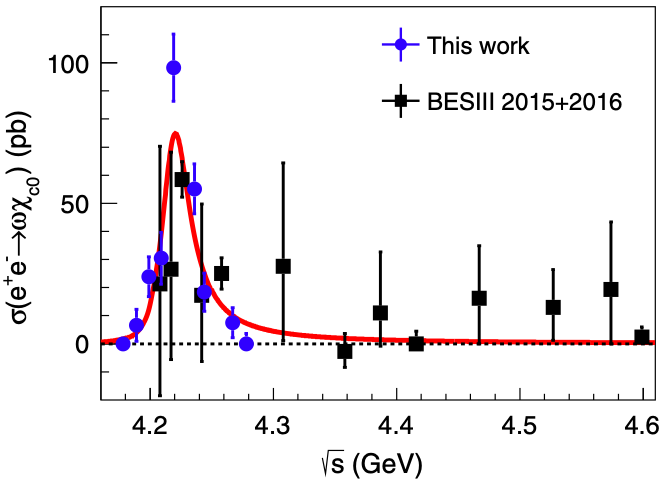}
	\vspace*{-0.3cm}
	\caption{The $e^{+}e^{-}\to\omega\chi_{c0}$ cross section as a function of the center-of-mass energy.
		The blue points are from Ref.~\cite{Ablikim:2019apl},
		the black square points are from Refs.~\cite{omegachic0-2015, omegachic0-2016},
		and the red solid line is the fit result.}
	%\vspace*{-0.6cm}
	\label{fig-omegachic0-cs}
\end{figure}

\begin{figure}[hbt]
  \centering
  \includegraphics[width = 0.40\textwidth]{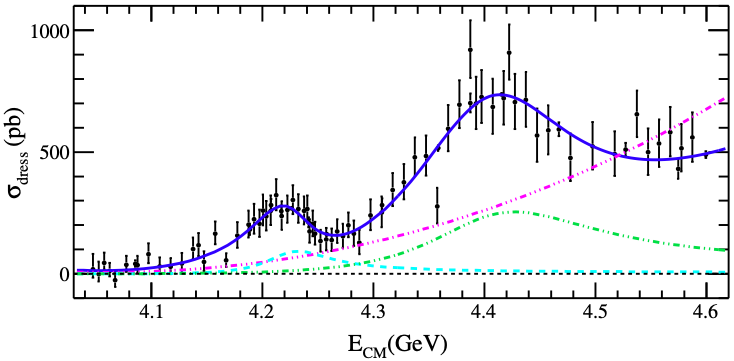}
  \vspace*{-0.3cm}
  \caption{Fit to the dressed cross sections of $e^{+}e^{-}\to \pi^{+} D^{0}D^{*-}$, 
		where the black dots with error bars are the measured cross sections and the blue line shows the fit result. 
		The error bars are statistical only. The pink dashed triple-dot line describes the phase-space contribution, 
		the green dashed double-dot line describes the recond resonance contribution, and the light blue dashed line describes the first resonance contribution.}
  %\vspace*{-0.6cm}
  \label{fig-pipD0Dstarm-cs}
\end{figure}

\begin{figure}[hbt]
  \centering
  \includegraphics[width = 0.25\textwidth]{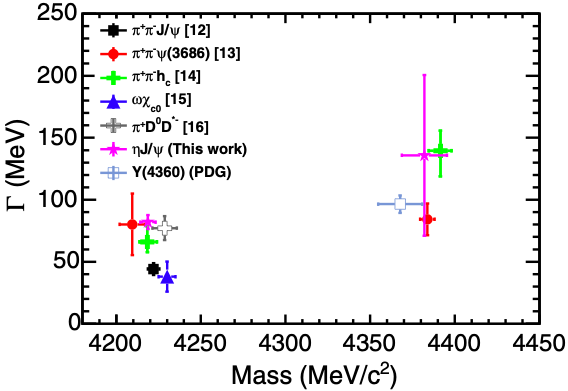}
  \vspace*{-0.3cm}
  \caption{Masses versus widths of the $Y(4220)$ and $Y(4360)$ obtained from the different final states by BESIII~\cite{etajpsi-1} 
		and $Y(4360)$ quoted from PDG~\cite{pdg}. Here the errors reflect both statistical and systematical uncertainties.}
  %\vspace*{-0.6cm}
  \label{fig-cmp-y4220}
\end{figure}

 Recently, BESIII experiment performed a lot of studies about $Y(4220)$ via different decay modes,
such as $e^{+}e^{-}\to\omega\chi_{c0}$~\cite{Ablikim:2019apl}, $e^{+}e^{-}\to \pi^{+} D^{0}D^{*-}$~\cite{Ablikim:2018vxx}, 
and $e^{+}e^{-}\to \eta J/\psi$~\cite{Ablikim:2020cyd}. 
From the cross section distributions of $e^{+}e^{-}\to \omega \chi_{c0}$ (Fig.~\ref{fig-omegachic0-cs}),
$e^{+}e^{-}\to \pi^{+} D^{0}D^{*-}$ (Fig.~\ref{fig-pipD0Dstarm-cs}), and $e^{+}e^{-}\to \eta J/\psi$ (Fig.~\ref{fig-etajpsi-cs}),
a common significant structure can be found around $\sqrt{s} = 4.22$ GeV, which is referred as $Y(4220)$.
To extracted the mass and width of the structure from those cross section distributions,
the Breit-Wigner, defined as Eq.~\ref{eq-bw}
\begin{equation}
	\label{eq-bw}
	BW = \frac{M}{\sqrt{s}}\frac{\sqrt{12\pi\Gamma_{ee}\Gamma Br}}{s - M^{2} +i M \Gamma} \sqrt{\frac{\Phi(\sqrt{s})}{\Phi(M)}},
\end{equation}
where $M$, $\Gamma$ are the mass and width of the structure,
while $\Gamma_{ee}$, $Br$, and $\Phi$ are the electric width of the structure, branching fraction of corresponding decay mode, 
and phase space (PHSP) factor, respectively, 
is used. 
The interference and continuum are considered if more structures.
The determined mass and width of $Y(4220)$ based on different decays are shown in Fig.~\ref{fig-cmp-y4220}, as well as $Y(4360)$.
For $Y(4220)$ and $Y(4360)$, the mass is compatible, while the width is not compatible very well.

\begin{figure}[hbt]
  \centering
  \includegraphics[width = 0.25\textwidth]{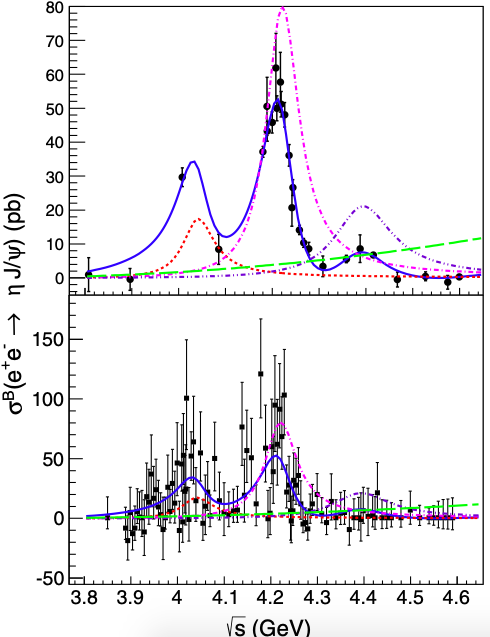}
  \vspace*{-0.3cm}
  \caption{Top: Cross section and fits of $e^{+}e^{-}\to \eta J/\psi$ for $XYZ$
		data. Bottom: Same for the scan data. Dots with error bars are data. 
		The solid (blue) curves represent the fit results of the following interfering amplitudes:
		$Y(4040)$ (dashed red), $Y(4220)$ (short-dashed pink), $Y(4360)$ (short-dashed purple), and P-PHSP (long-dashed green).}
  %\vspace*{-0.6cm}
  \label{fig-etajpsi-cs}
\end{figure}

\subsection{$e^{+}e^{-}\to\pi^{0}\pi^{0}J/\psi$}
\par The Born cross sections of the process $e^{+}e^{-}\to\pi^{0}\pi^{0}J/\psi$ at center-of-mass 
energies between 3.808 and 4.600 GeV are measured with high precision by using 12.4 fb$^{-1}$ of data samples at BESIII~\cite{BESIII:2020pov}.
To describe the distribution, a $\chi^{2}$ fit is performed with $Y(4220)$, $Y(4320)$ and nonresonant component.
The resonance is described by Eq.~\ref{eq-bw}.
Here, the mass and width of $Y(4320)$ are fixed as those taken from Ref.~\cite{Ablikim:2016qzw}, while those of $Y(4220)$ float.
The nonresonant component in the Born cross sections is decribed as $\Phi(\sqrt{s}) e^{-p_{0}(\sqrt{s} - 2m_{\pi^{0}} - m_{J/\psi}) + p_{1}}$,
where $p_{0}$ and $p_{1}$ are the float parameters.
The mass and width of the $Y(4220)$ are fitted to be $(4220.4 \pm 2.4 \pm 2.3)$ MeV/$c^{2}$ and $(46.2 \pm 4.7 \pm 2.1)$ MeV, respectively, 
where the first uncertainties are statistical and the second systematic.
The statistical significance of $Y(4320)$ is estimated to be 4.2$\sigma$ by changes in the $\chi^{2}$ and ndf values obtained from 
including and excluding the $Y(4320)$.

\begin{figure}[hbt]
	\centering
	\includegraphics[width = 0.25\textwidth]{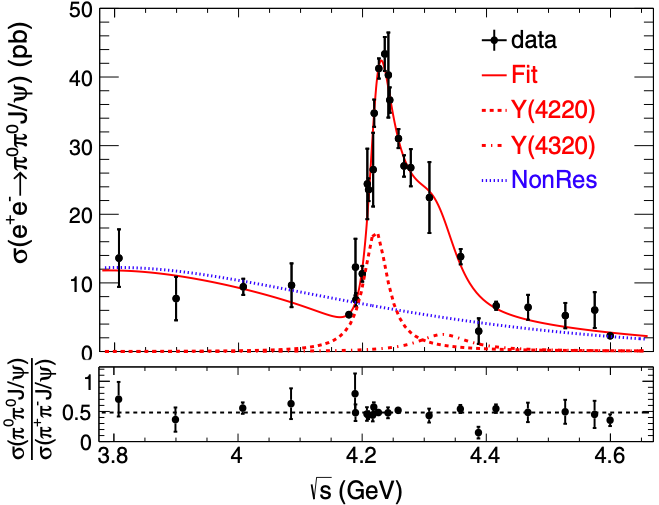}
	\vspace*{-0.3cm}
  \caption{Top: Fit to the Bron cross sections of $e^{+}e^{-}\to\pi^{0}\pi^{0}J/\psi$,
		where points with error bars are data, the red solid line is the total fit result, 
		the blue dotted line is the nonresonant component, while the red dashed and dot-dashed lines represent the contributions 
		from $Y(4220)$ and $Y(4320)$, respectively. 
		Bottom: Cross-section ratio of $e^{+}e^{-} \to \pi^{0} \pi^{0} J/\psi$ to $e^{+}e^{-} \to \pi^{+} \pi^{-} J/\psi$, 
		where the black dashed line corresponds to the average.}
  %\vspace*{-0.6cm}
  \label{fig-pizpizjpsi-cs}
\end{figure}

\subsection{$e^{+}e^{-}\to\mu^{+}\mu^{-}$}
\par
\begin{figure}[hbt]
  \centering
  \includegraphics[width = 0.25\textwidth]{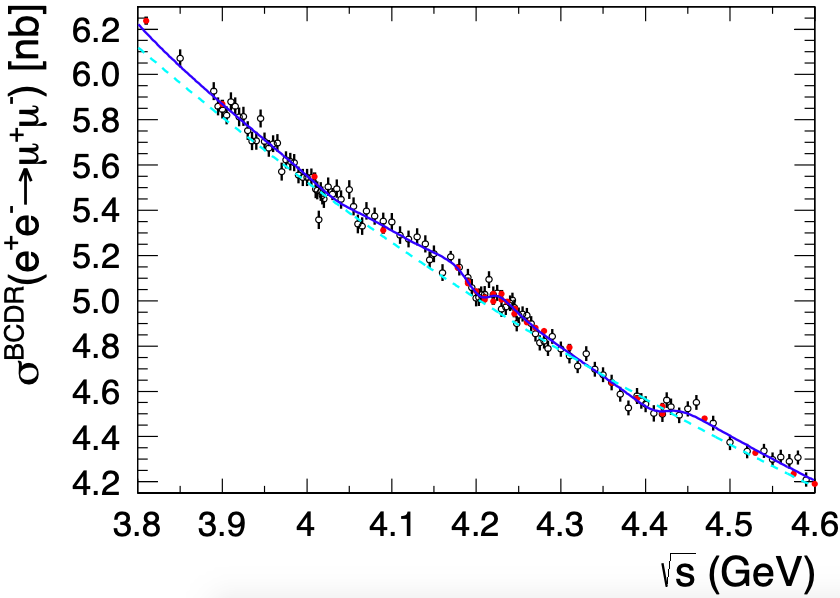}
  \includegraphics[width = 0.25\textwidth]{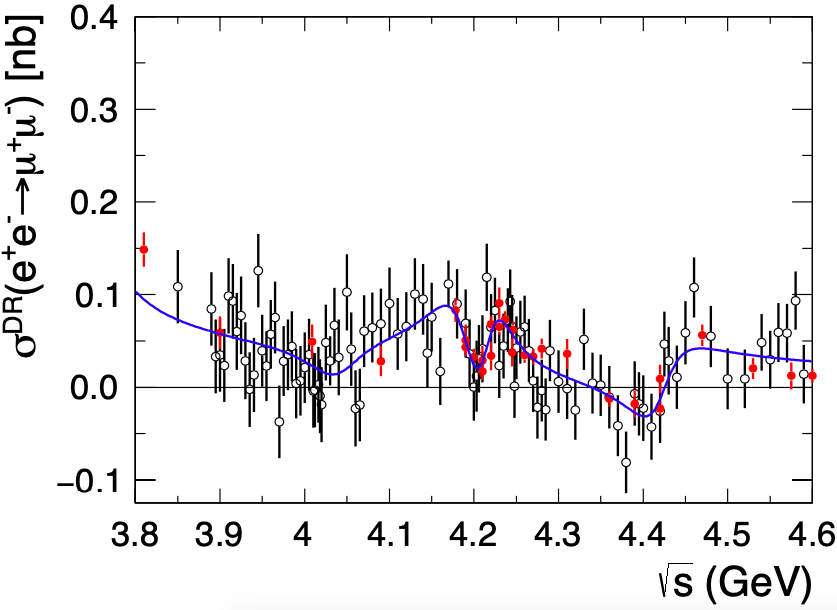}
  \vspace*{-0.3cm}
  \caption{Measured cross sections for $e^{+} e^{-} \to \mu^{+} \mu^{-}$ with the fit superimpose. 
					The top plot shows the absolute cross sections, while the bottom plot shows the cross sections after subtraction of both the continuum and 
					$\psi(3686) \to \mu^{+} \mu^{-}$ contributions}
  %\vspace*{-0.6cm}
  \label{fig-mumui-cs}
\end{figure}

 Using data sample with $\sqrt{s} = 3.8 - 4.6$ GeV, the widths and the branching fractions of $Y(4040)$, $Y(4160)$, and $Y(4415)$
decaying to $\mu^{+}\mu^{-}$ are measured, and the phase of the amplitudes are determined~\cite{BESIII:2020peo}. 
The measured muonic widths for the $Y(4040)$ and
$Y(4415)$ are consistent within $\sim 1.3\sigma$ with theoretical expectations for the electronic width of these states,
which are 1.42 keV for $Y(4040)$ and 0.70 keV for $Y(4415)$~\cite{Li:2009zu}.
To describe the cross sections of $e^{+}e^{-}\to \mu^{+}\mu^{-}$, coherent fit with conponents of continuum and ten resonances
($\rho$, $\omega$, $\phi(1020)$, $J/\psi$, $\psi(3686)$, $\psi(3770)$, $Y(4040)$, $Y(4160)$, $Y(4415)$ and $Y(4220)$) are performed. 
The amplitude for continuum $e^{+}e^{-}\to \mu^{+}\mu^{-}$ production is taken from Ref.~\cite{mumu-cont1, mumu-cont2},
and resonance is described with BreitWigners shown in Eq.~\ref{eq-bw}.
The top one of Fig.~\ref{fig-mumui-cs} shows fit result of the sum of the Born continuum cross sections 
of $e^{+}e^{-}\to \mu^{+}\mu^{-}$ and the dressed cross sections for the reconances
decaying into $\mu^{+}\mu^{-}$, while the bottom one shows that of the continuum
and $\psi(3686) \to \mu^{+} \mu^{-}$ contributions subtracted 
dressed cross sections. 
All the cross sections in Fig.~\ref{fig-mumui-cs} have corrected the total dimuon cross sections for radiative effects. 
The evidence for structure $Y(4220)$ is found with mass $M = 4216.7 \pm 8.9(stat) \pm 4.1(sys)$ MeV/$c^{2}$ 
and width $\Gamma = 47.2 \pm 22.8(stat) \pm 10.5(sys)$ MeV.
The statistical significance of $Y(4220)$ is more than 7$\sigma$ if taken into consideration radiative effects.

\subsection{Other searches}
\par
\begin{figure}[hbt]
	\centering
	\includegraphics[width = 0.25\textwidth]{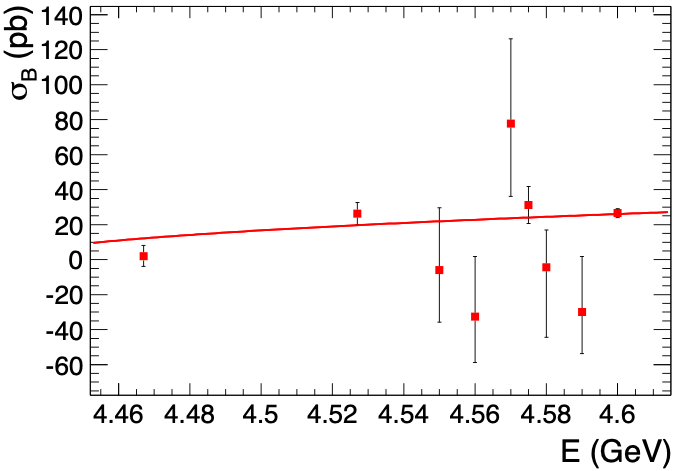}
	\includegraphics[width = 0.25\textwidth]{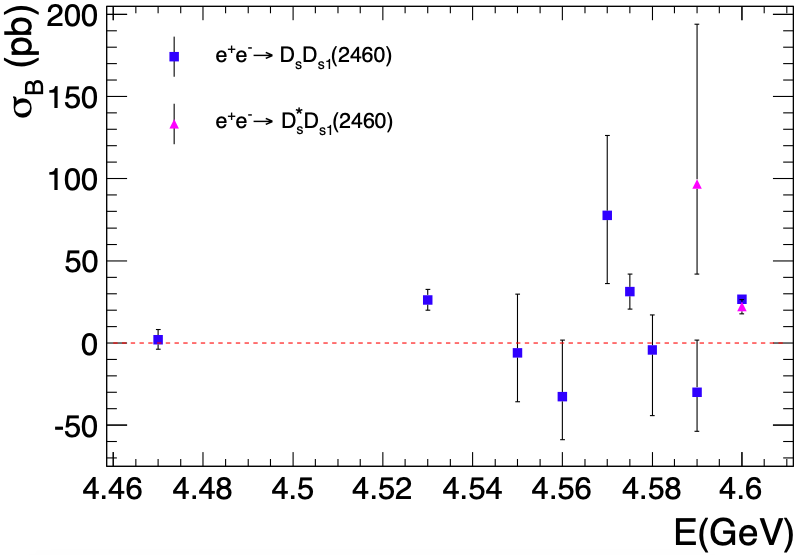}
	\vspace*{-0.3cm}
  \caption{Fit of Born cross sections of $e^{+}e^{-} \to D_{s}^{+} D_{s1}(2460)^{-}$ with $\sigma^{B} \propto \sqrt{E_{c.m.} - E_{0}}$ (top), and
	comparison of the Born cross section of $e^{+}e^{-} \to D_{s}^{+} D_{s1}(2460)^{-}$ and $e^{+}e^{-} \to D_{s}^{*+} D_{s1}(2460)^{-}$ (bottom).}
  %\vspace*{-0.6cm}
  \label{fig-dd2460-cs}
\end{figure}

 To study more about $Y$ states, BESIII tried a lot of other cross section measurements recently, 
such as $e^{+}e^{-} \to D_{s}^{+} D_{s1}(2460)^{-} +c.c.$ and $e^{+}e^{-} \to D_{s}^{*+} D_{s1}(2460)^{-} +c.c.$~\cite{Ablikim:2020cbi},
$e^{+} e^{-} \to \eta^{\prime} J/\psi$~\cite{Ablikim:2019bwn},  
$e^{+} e^{-} \to D^{+} D^{-} \pi^{+} \pi^{-}$~\cite{Ablikim:2019okg}.
For $e^{+}e^{-} \to D_{s}^{+} D_{s1}(2460)^{-}$ and $e^{+}e^{-} \to D_{s}^{+} D_{s1}(2460)^{-}$,
we can hardly find evidence of $Y$ state from the measured Born cross sections shown in Fig.~\ref{fig-dd2460-cs}.
For $e^{+}e^{-} \to \eta J/\psi$, 
we tried coherent fit of Born cross section of $e^{+}e^{-} \to \eta J/\psi$ with mass and width fixed $Y(4160)$ and $Y(4260)$.
For $e^{+}e^{-} \to D^{-} D^{+}_{1} + c.c.$ and $e^{+}e^{-} \to D^{-} D^{+}_{1} + c.c.$,
we measured the Born cross sections with partial reconstruction method, and find 
some indications of enhanced cross sections for both processes.

\begin{figure}[hbt]
	\centering
	\includegraphics[width = 0.25\textwidth]{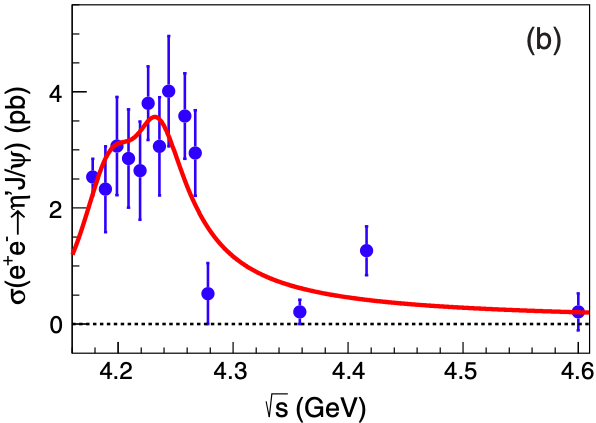}
	\vspace*{-0.3cm}
	\caption{Fit of Born cross sections of $e^{+}e^{-} \to \eta J/\psi$ with mass and width fixed $Y(4160)$ and $Y(4260)$.}
	\label{fig-etajpsi-cs}
\end{figure}

\begin{figure}[hbt]
	\centering
	\includegraphics[width = 0.25\textwidth]{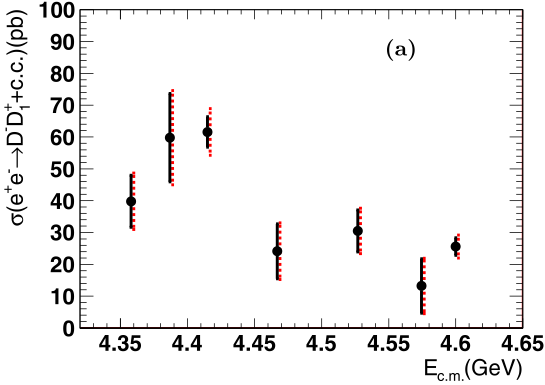}
	\includegraphics[width = 0.25\textwidth]{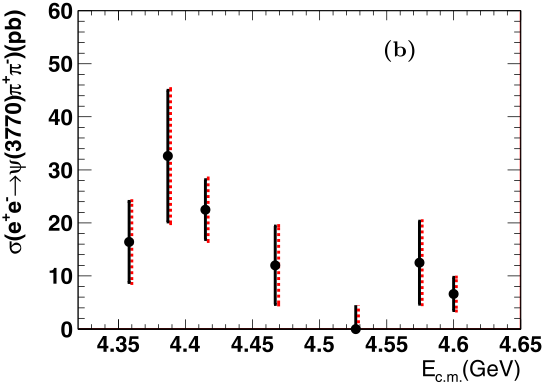}
	\vspace*{-0.3cm}
	\caption{The measured Born cross sections of the signal processes (a) $e^{+}e^{-} \to D_{1}(2420)^{+}D^{-}+c.c.\to D^{+}D^{-}\pi^{+}\pi^{-}$ and 
		(b) $e^{+}e^{-} \to \pi^{+}\pi^{-}\psi(3770) \to D^{+}D^{-}\pi^{+}\pi^{-}$. The (black) solid lines are the sum of statistical uncertainties 
		and independent systematic uncertainties in quadrature, the (red) dot lines are total uncertainties.}
	\label{fig-dd1-cs}
\end{figure}

\section{$Z$ states}
\begin{figure}[hbt]
	\centering
	\includegraphics[width = 0.45\textwidth]{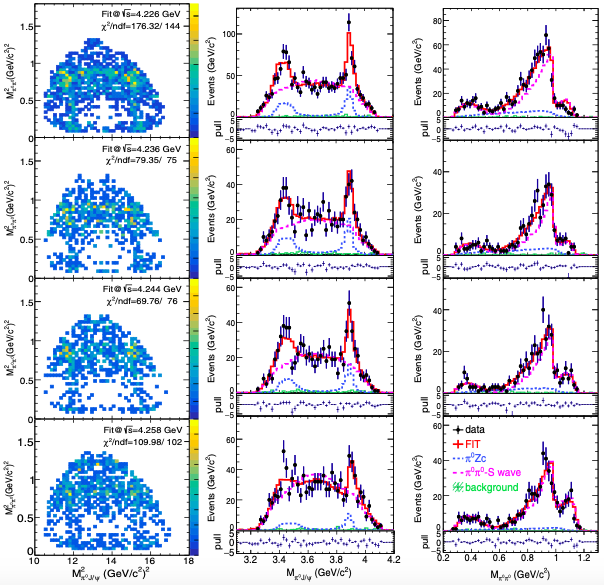}
	\vspace*{-0.3cm}
  \caption{(Left column) Dalitz plots of $M(\pi^{0}J/\psi)$ versus $M(\pi^{0}\pi^{0})$, invariant-mass projections (middle column) $M(\pi^{0}J/\psi)$
		and (right column) $M(\pi^{0}\pi^{0})$ of the results of the nominal PWA for data samples $\sqrt{s} = 4.226 - 4.258$ GeV. Points with errors
		are data, red solid curves are the total fit results, the blue dashed (magenta long-dashed) curves represent $Z_{c}(3900)^{0}$ ($\pi\pi$ S-wave) components, 
		and green shaded histograms represent the estimated backgrounds. Each event appears twice in the Dalitz plots and $M(\pi^{0}J/\psi)$ distributions. 
		The $\chi^{2}$/ndf is calculated by merging those bins with less than 10 events in the Dalitz plots.}
	\vspace*{-0.5cm}
  \label{fig-pwa-mpijpsi}
\end{figure}

The neutral $Z_{c}(3900)^{0}$ was observed in the processes $e^{+}e^{-} \to \pi^{0}\pi^{0} J/\psi$ and $e^{+} e^{-} \to \pi^{0} (D\bar{D}^{*})^{0}$
by CLEO-c and BESIII~\cite{zc0-cleo, zc0-bes3-1, zc0-bes3-2}.
To investigate more about $Z_{c}(3900)^{0}$ properties, using 12.4 fb$^{-1}$ data samples, 
an amplitude analysis of $e^{+}e^{-} \to \pi^{0}\pi^{0} J/\psi$ is performed~\cite{BESIII:2020pov}.
In the nominal fit, the intermediate processes $e^{+} e^{+} \to \sigma J/\psi$, $f_{0}(980)J/\psi$,
$f_{0}(1370) J/\psi$, and $\pi^{0} Z_{c}(3900)^{0}$ are included. 
Based on the simultaneous partial wave analysis fit results, 
as shown in Fig.~\ref{fig-pwa-mpijpsi}, the $\pi^{0}\pi^{0}$ S-wave contribution dominates.
The spin-parity of $Z_{c}(3900)^{0}$ is determined to be $J^{P} = 1^{+}$ with a statistical significance of more than 9$\sigma$,
and the mass and width are measured to be $(3893.0 \pm 2.3 \pm 19.9)$ MeV/$c^{2}$ and $(44.2 \pm 5.4 \pm 9.1)$ MeV, respectively.
These values are consistent with those of the charged $Z_{c}(3900)^{\pm}$ observed in $e^{+}e^{-} \to \pi^{+} \pi^{-} J/\psi$.

\section{$X$ states}

\begin{figure}[!h]
	\centering
	\includegraphics[width = 0.40\textwidth]{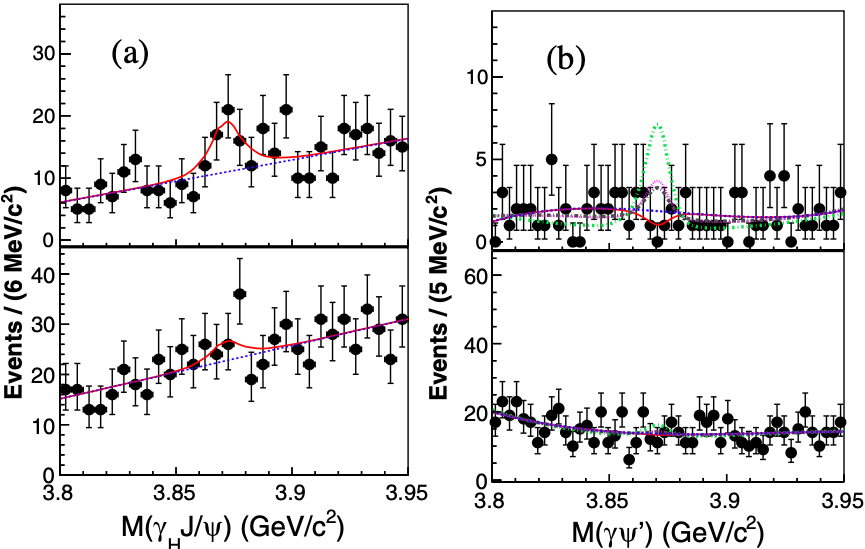}
	\vspace*{-0.3cm}
	\caption{
		(a) Fit results for $X(3872) \to \gamma J/\psi$ for the $\mu^{+} \mu^{-}$ (top) and $e^{+} e^{-}$ (bottom) mode. 
		(b) Fit results for $X(3872) \to \gamma \psi(3686)$ for the $\pi^{+}\pi^{-}J/\psi$ (top) and $\mu^{+}\mu^{-}$ (bottom) mode. 
		The points with error bars are from data, the red curves are the best fit. 
		In (b), the rose-red dotted line represents the fit with the signal constrained to the expectation using $X(3872)\to\pi^{+}\pi^{-}J/\psi$ 
		based on the relative ratios taken from a global fit~\cite{Li:2019kpj}; 
		the green dash-dotted lines are using $X(3872) \to \gamma J/\psi$ as the reference based on the LHCb measurement~\cite{Aaij:2014ala}, 
		and the gray long dashed lines 
		are using $X(3872) \to \gamma J/\psi$ as the reference based on the Belle measurement~\cite{Bhardwaj:2011dj}.
	}
	\label{fig-x3872-gamjpsi}
\end{figure}

\begin{figure}[!h]
	\centering
	\includegraphics[width = 0.40\textwidth]{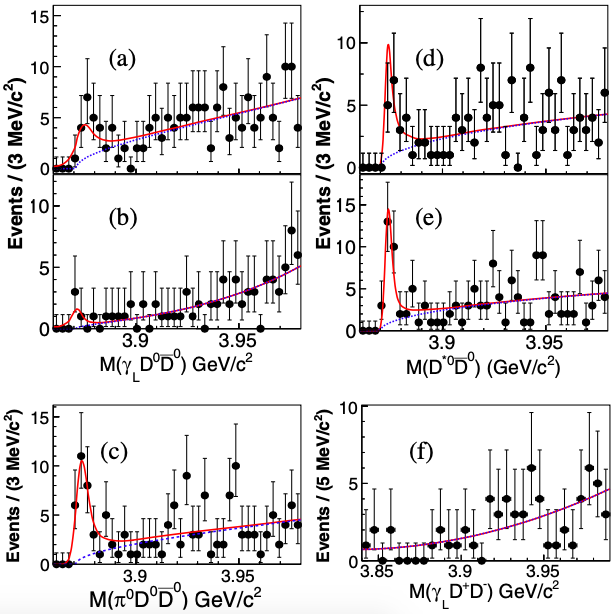}
	\vspace*{-0.3cm}
  \caption{
		$M(\gamma_{L} D^{0} \bar{D}^{0})$ with $M(\gamma_{L} D^{0})$ (a) in or (b) below the $D^{*0}$
		mass window. (c) $M(\pi^{0}D^{0}\bar{D}^{0})$ with $M(\pi^{0}D^{0})$ in the $D^{*0}$ mass
		window. Simultaneous fit results for $X(3872) \to D^{*0}\bar{D}^{0}$ with
		(d) $D^{*0} \to \gamma D^{0}$ and (e) $D^{*0} \to \pi^{0}D^{0}$ mode. (f) Fit results for $X(3872) \to \gamma_{L}D^{+}D^{-}$. 
		The points with error bars are from data, the red curves are the best fit, and the blue dashed curves are the background components.
	}
	\label{fig-x3872-dd}
\end{figure}

\par Using data at $\sqrt{s} = 4.178 - 4.278$ GeV, BESIII tried to search for new decay modes for $X(3872)$~\cite{Ablikim:2020xpq}.
We succeed to observe $X(3872) \to D^{*0}\bar{D}^{0} + c.c.$ and and find evidence for $X(3872) \to \gamma J/\psi$ 
with significances of $7.4\sigma$ and $3.5\sigma$, respectively,
but failed to find evidence for  $X(3872) \to \gamma \psi(3686)$ and $X(3872) \to \gamma D^{+}D^{-}$.
We extract the signals based on unbinned maximum likelihood fit as shown in Fig.~\ref{fig-x3872-dd} and Fig.~\ref{fig-x3872-gamjpsi},
and calculate the branching fractions or upper limits.
The upper limit of the ratio $Br(X(3872) \to \gamma \psi(3686)) / Br(X(3872) \to \gamma J/\psi)$ 
is determined as $<0.59$ at 90\% C.L., which is consistent with the Belle measurement~\cite{Bhardwaj:2011dj} and the global fit~\cite{Li:2019kpj},
but challenges the LHCb measurement~\cite{Aaij:2014ala}.
This measurement, taking into account model predictions, suggests that the $X(3872)$ state is more likely a molecule or a mixture of molecule and charmonium,
rather than a pure charmonium state. Also, this study provides essential input to future tests of the molecular model for the $X(3872)$ state.

\section{Summary}
\par BESIII experiment has achieved a lot of progresses on $XYZ$ states, recently, especially for $Y(4220)$, $Y(4360)$, $Z_{c}(3900)^{0}$, and $X(3872)$.
In 2019-2020, BESIII experiment collected more $XYZ$ data samples at $\sqrt{s} = 4.620 - 4.700$ GeV, which support us to do furhter studies about $XYZ$ states.
Besides, BESIII experiment plan to collect new data above $\sqrt{s} = 4.700$ GeV. We believe more interesting analysis on $XYZ$ will coming.

%%%%%%%%%%%%%%%%%%%%%  %%%%%%%%%%%%%%%%%%%%


\begin{thebibliography}{999}
\bibitem{etajpsi-1} For $\eta J/\psi$ process, the PHSP, $Y(4040)$, $Y(4220)$ and $Y(4360)$ components are considered. 
		For $\pi^{+}\pi^{-} J/\psi$ process, the exponential continuum, $\psi(3770)$, $Y(4220)$ and $Y(4230)$ components are considered. 
		For $\pi^{+}\pi^{-}\psi(3686)$ process, the $Y(4220)$ and $Y(4360)$ components are considered. 
		For $\pi^{+}\pi^{-}h_{c}$ process, the $Y(4220)$ and $Y(4360)$ components are considered. 
		For $\pi^{+}D^{0}D^{*-}$ process, The PHSP, $Y(4220)$ and $Y(4360)$ components are considered.
\end{thebibliography}
\end{document}